# Purification of molybdenum oxide, growth and characterization of medium size zinc molybdate crystals for the LUMINEU program [a]

V.N. Shlegel[1,b], L. Berge[2], R.S. Boiko[3], M. Chapellier[2], D.M. Chernyak[2,3], N. Coron[4], F.A. Danevich[3], R. Decourt[5], V.Ya. Degoda[6], L. Devoyon[7], A. Drillien[2], L. Dumoulin[2], C. Enss[8], A. Fleischmann[8], L. Gastaldo[8], A. Giuliani[2], M. Gros[7], S. Herve[7], I.M. Ivanov[1], V.V. Kobychev[3], Ya.P. Kogut[6], F. Koskas[7], M. Loidl[7], P. Magnier[7], E.P. Makarov[1], M. Mancuso[2,9], P. de Marcillac[4], S. Marnieros[2], C. Marrache-Kikuchi[2], S.G. Nasonov[1], X.F. Navick[7], C. Nones[7], E. Olivieri[2], B. Paul[7], Y. Penichot[7], G. Pessina[10], O. Plantevin[2], D.V. Poda[2,3], T. Redon[4], M. Rodrigues[7], O. Strazzer[7], M. Tenconi[2], L. Torres[4], V.I. Tretyak[3], Ya.V. Vasiliev[1], M. Velazquez[5], O. Viraphong[5] and V.N. Zhdankov[11]

[1] Nikolaev Institute of Inorganic Chemistry, 630090 Novosibirsk, Russia
[2] Centre de Sciences Nucléaires et de Sciences de la Matière, 91405 Orsay, France
[3] Institute for Nuclear Research, MSP 03680 Kyiv, Ukraine
[4] IAS, Bâtiment 121, UMR 8617 Université Paris-Sud 11/CNRS, 91405 Orsay, France
[5] CNRS, Université de Bordeaux, ICMCB, 87 avenue du Dr. A. Schweitzer, 33608 Pessac cedex, France
[6] Kyiv National Taras Shevchenko University, MSP 03680 Kyiv, Ukraine
[7] CEA-Saclay, F-91191 Gif sur Yvette, France
[8] Institut für Angewandte Physik, Universität Heidelberg, Albert-Ueberle-Strasse 3-5, D-69120 Heidelberg, Germany
[9] Dipartimento di Scienza e Alta Tecnologia dell'Università dell'Insubria, I-22100 Como, Italy
[10] Dipartimento di Fisica dell'Università di Milano-Bicocca e Sezione di Milano Bicocca dell'INFN, Italy
[11] CML Ltd., 630090 Novosibirsk, Russia

**Abstract.** The LUMINEU program aims at performing a pilot experiment on neutrinoless double beta decay of $^{100}$Mo using radiopure ZnMoO$_4$ crystals operated as scintillating bolometers. Growth of high quality radiopure crystals is a complex task, since there are no commercially available molybdenum compounds with the required levels of purity and radioactive contamination. This paper discusses approaches to purify molybdenum and synthesize compound for high quality radiopure ZnMoO$_4$ crystal growth. A combination of a double sublimation (with addition of zinc molybdate) with subsequent recrystallization in aqueous solutions (using zinc molybdate as a collector) was used. Zinc molybdate crystals up to 1.5 kg were grown by the low-thermal-gradient Czochralski technique, their optical, luminescent, diamagnetic, thermal and bolometric properties were tested.

## 1 Introduction

Neutrinoless double beta (0ν2β) decay is a key process in particle physics thanks to its unique ability to test the Majorana nature of neutrino and lepton number conservation, the absolute scale and the hierarchy of neutrino mass [1, 2, 3, 4]. Low temperature scintillating bolometers are considered as extremely promising detectors to search for 0ν2β decay in different nuclei [5, 6, 7, 8, 9, 10, 11]. Recently developed technique to grow large high quality radiopure zinc molybdate (ZnMoO$_4$) crystal scintillators [9, 10, 12, 13, 14, 15] makes this material advantageous for low temperature bolometric experiments to search for 0ν2β decay of $^{100}$Mo. Here we report further progress in deep purification of molybdenum and growth of ZnMoO$_4$ crystals for the LUMINEU project. First results of the crystals characterization are presented too.

## 2 Production of ZnMoO$_4$ crystals

### 2.1 Purification of molybdenum

High purity molybdenum and zinc are required to grow high quality radiopure ZnMoO$_4$ crystal scintillators. While a high purity zinc oxide is commercially available, molybdenum should be additionally purified. Furthermore, there are no commercially available molybdenum compounds that are tested for the presence of radioactive elements and have the required level of radioactive contamination. Such a test of the raw materials for crystal growth is extremely difficult and

---

requires long measurement procedure. Typically high sensitivity radiopurity tests can be only done after crystal growth using calorimetric method. Moreover, development of efficient purification methods with minimal losses of molybdenum is strongly required to produce $ZnMoO_4$ crystals from enriched molybdenum, whose contamination is typically on the level of tens – hundreds ppm [16] (keeping also in mind the necessity to recycle the costly enriched material). We have developed a two stages technique of molybdenum purification consisting of sublimation of molybdenum oxide in vacuum (with addition of zinc molybdate) and double recrystallization from aqueous solutions by co-precipitation of impurities on zinc molybdate sediment.

### 2.1.1 Purification of $MoO_3$ by sublimation

Sublimation of molybdenum oxide under atmospheric pressure with subsequent leaching in aqueous solutions with ammonia is widely used in the industry of molybdenum. Nevertheless the concentration of impurities, particularly of tungsten (on the level of up to 0.5wt% even in the high purity grade materials) still exceeds the $ZnMoO_4$ crystal growth requirements. Even additional vacuum sublimation of molybdenum oxide proved to be insufficient. According to [17] separation of tungsten and molybdenum is a well known problem. Besides, the sublimation of $MoO_3$ is not efficient enough to reduce traces e.g. of Ca, Na and Si to the level below 20 – 70 ppm.

We have assumed that during sublimation at high temperature the following exchange reaction could occur:

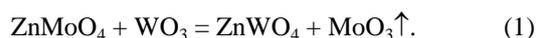

$$ZnMoO_4 + WO_3 = ZnWO_4 + MoO_3\uparrow. \qquad (1)$$

Such a reaction should reduce the concentration of tungsten, and therefore can be used for separation of molybdenum from tungsten. To prove this possibility, we have prepared a sample of $MoO_3$ powder with 10wt% of $WO_3$. The concentration of tungsten in the $MoO_3$ product after sublimation was reduced to 0.1wt%.

One more confirmation of the method's efficiency was obtained by chemical and X-ray diffraction analysis of the rests after the sublimations performed with an aim to purify molybdenum for crystal growth (the amount of the rests is typically 1 – 3 wt% of the initial amount of the purified material). The bottoms after a few sublimation processes were mixed and annealed in the air atmosphere to oxidize residues of metals. Then we have carried out sublimation of the sample in vacuum to reduce presence of $MoO_3$. Atomic emission analysis, performed in the analytical laboratory of the Nikolaev Institute of Inorganic Chemistry, gives the following elemental composition of the bottoms: Ca – 0.14wt%; Cu – 0.011wt%; Fe – 0.064wt%; K – 1wt%; Mg – 0.026wt%; Na – 0.13wt%; Si – 2.6wt%; Mo – 22wt%; W – 18wt%; Zn – 14wt%. Oxides of molybdenum and silicon, tungstate (in form of tungstate-molybdate) and molybdate of zinc, as well as $K_2Mo_7O_{22}$ and $K_2MgSi_5O_{12}$, have been identified in the bottoms with the help of X-ray diffraction analysis. At the same time, tungsten oxide (present in the initial product) was not detected in the bottoms. The data supported occurring of the exchange reaction (1) and confirmed efficiency of molybdenum oxide sublimation in vacuum with addition of zinc molybdate.

To purify molybdenum for $ZnMoO_4$ crystal growth, we have added up to 1% of high purity zinc molybdenum (obtained earlier in the course of the R&D) to the $MoO_3$ prepared for sublimation. The obtained sublimates contained mixture of molybdenum oxides of different composition and color, which hinders their use for $ZnMoO_4$ synthesis. The sublimates were then annealed in the air atmosphere to obtain yellow color stoichiometric $MoO_3$. The sublimates were analyzed by atomic emission spectrometry. The results are presented in Table 1. One can see that the purity level of $MoO_3$ was improved one–two orders of magnitude after the double sublimation process. The sublimation also should remove metal oxides, which have a high vapor pressure at temperatures up to a thousand degrees.

**Table 1.** Efficiency of molybdenum oxide purification by sublimation.

| Material | Concentration of impurities (ppm) | | | |
|---|---|---|---|---|
| | Si | K | Fe | W |
| Initial $MoO_3$ | 600 | 100 – 500 | 6 | 200 – 500 |
| After 1st sublimation | 100 – 500 | 10 – 50 | 2 – 6 | 100 – 200 |
| After 2nd sublimation | 70 | 1 – 8 | < 1 | 30 – 40 |

### 2.1.2 Purification by recrystallization from aqueous solutions

Finally the molybdenum was purified by double recrystallization of ammonium molybdate in aqueous solutions with the deposition of impurities on zinc molybdate sediment. For this purpose molybdenum oxide was dissolved in solution of ammonia. Mono-molybdates and various poly-compounds and hetero-poly compounds are formed depending on the mixing ratio of the components. A composition of the compounds depends on the acidity of the solution and components concentration. Molybdates in aqueous solutions form normal molybdate:

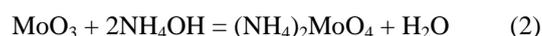

$$MoO_3 + 2NH_4OH = (NH_4)_2MoO_4 + H_2O \qquad (2)$$

and ammonium hepta-molybdates, $(NH_4)_xH_{(6-x)}[Mo_7O_{24}]$. Poly-molybdates can form, in presence of impurities, during long time exposure, ammonium salts of silico-molybdic and phosphor-molybdic acids, for instance $H_8[Si(Mo_2O_7)_6]$, $H_7[P(Mo_2O_7)_6]$. Role of the central atom, except Si(IV) and P(V), can perform also V(V), Ge(IV), Cr(III), etc., while ligands of the inner sphere can be ions $WO_4^{2-}$, $VO^{3-}$, $CrO_4^{2-}$, $TeO_4^{2-}$. Solutions of poly-molybdates at pH < 6 are able to dissolve oxides and hydroxides of many metals, e.g. ZnO, $Fe(OH)_3$, $Ni(OH)_2$, $Cu(OH)_2$, etc. As a result, a partial co-crystallization of impurities could occur. Thus, recrystallization of ammo-nium para-molybdate is not efficient enough for molybdenum purification. Besides, a typical concentration of impurities in high purity commercial MoO$_3$ is relatively low (1 – 100 ppm). As a result, the impurity sediments appear in a form of fine microcrystals, hardly removable by filtration.

**Table 2.** Purity level of MoO$_3$ before and after purification by recrystallization and sublimation. Data for commercial 5N5 grade product, and enriched $^{100}$Mo material are given for comparison.

| Material | Concentration of impurities (ppm) | | | | | | | |
|---|---|---|---|---|---|---|---|---|
| | Na | Mg | Si | K | Ca | Fe | Zn | W |
| Initial MoO$_3$ | 60 | 1 | 60 | 50 | 60 | 8 | 10 | 200 |
| Recrystallization from aqueous solutions | 30 | < 1 | 30 | 20 | 40 | 6 | 1000 | 220 |
| Sublimation and recrystallization from aqueous solutions | - | < 1 | 30 | 10 | 12 | 5 | 500 | 130 |
| Double sublimation and recrystallization from aqueous solutions | - | < 1 | - | < 10 | < 10 | < 5 | 70 | < 50 |
| 5N5 grade MoO$_3$ used to produce ZnMoO$_4$ crystal studied in [12, 26] | 24 | - | 9 | 67 | 15 | <18 | - | 96 |
| Samples of enriched isotope $^{100}$Mo used in [16] (before purification, data of producer) | 10 | < 10 | 50–360 | < 30 | 40-50 | 10-80 | - | 200 |

To improve efficiency of the recrystallization process, we have used zinc oxide to initiate precipitation (taking into account that zinc does not affect the crystal quality). ZnO on the level of 1 – 2 g/L was dissolved in the ammonium para-molybdate solution at pH > 6, then ammonia was added to the solution to reach pH = 7 – 8. After several hours of exposure precipitation of zinc molybdate occurs. The ZnMoO$_4$ sediment sorbs impurities from the solution. Further increasing of pH leads to precipitation of contaminants in the form of hydroxides. It should be stressed, the basic solution with pH ≈ 8 – 9 provides the most favorable conditions for thorium and uranium precipitation. After separation of the sediment, the solution was evaporated to 70%. Then ammonium oxalate was added to the solution to bind the residues of iron impurities. Results of the purification are presented in Table 2.

It should be also stressed that using of the additional "wet" chemistry procedure is also encouraged by the fact that large crystal grains of MoO$_3$ are formed in sublimation process, which provides certain difficulties to produce radiopure ZnMoO$_4$ powder (an additional procedure of the oxide grains grinding could contaminate the material). Subsequent dissolution of the molybdenum oxide in ammonia allows to obtain high purity MoO$_3$ perfectly fine for further synthesis of ZnMoO$_4$ powder.

The molybdenum oxide purified by twice recrystallization procedure from aqueous solutions and high purity grade ZnO produced by Umicore were used to synthesize ZnMoO$_4$ powder for crystal growth.

## 2.2 ZnMoO$_4$ crystal growth

Several ZnMoO$_4$ crystal boules were grown in air atmosphere from the purified input powder by the low-thermal-gradient Czochralski technique [18, 19, 20] in platinum crucibles ⌀40 and ⌀80 mm (it should be mentioned that, according to the certificates of the platinum crucibles, iron content in the platinum does not exceed 40 ppm). The temperature gradient was kept below 1 K/cm, the rotational speed was in the range of 5 – 20 rotations per minute with the crystallization rate of 0.8 – 1.2 mm/hour. A low crystallization rate was kept during growing of upper cone of the crystal boules. Rotational speed was decreased from the start to the end of the growth process in 1.5 – 2 times. The yield of the produced boules was on the level of 80%, which is an important achievement taking into account the future plans to produce crystals from enriched $^{100}$Mo. Four optical elements (two ⌀20 × 40 mm and two ⌀35 × 40 mm with masses 55 g and 160 g, respectively) were cut and polished for low temperature measurements. Several small samples were produced for optical, luminescent, diamagnetic and thermal tests.

## 3 Characterization of ZnMoO$_4$ crystals

### 3.1 Optical absorption

Visible and near infrared absorption spectra of ZnMoO$_4$ crystal were recorded with a Varian Cary 5000 spectrophotometer. The transmission coefficient, T, was measured on a 2.0 mm-thick single crystal and found to be higher than 0.5 from 327 nm to 4.96 μm. The absorption coefficient was calculated as $\alpha = -\log T \times \ln 10 / t$, with $t$ the thickness of the crystal. The data are presented in Fig. 1.

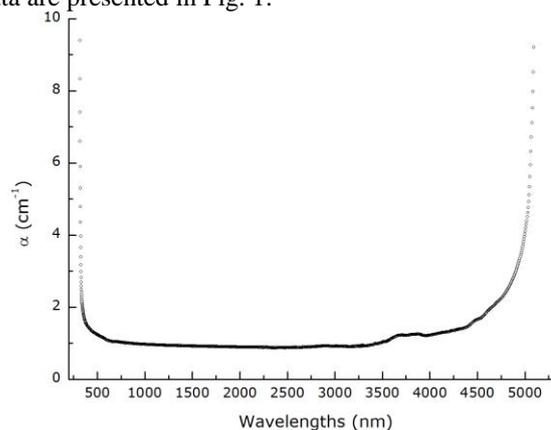

**Figure 1.** Absorption of 2 mm thick ZnMoO$_4$ single crystal. α denotes absorption coefficient.

The cut-off wavelengths are 313 nm ($\approx$ 3.96 eV) and 5.13 µm. The absorption coefficient decreases from 1.47 to 0.89 cm$^{-1}$ in the wavelength region from 400 nm to 2 µm, but there does not seem to have a broad absorption band around 440 nm that could be ascribed to $Fe^{2+}/Fe^{3+}$ impurities as described in [21] and [22]. This is due to the low Fe concentration in the crystal ($\approx 1.71 \times 10^{16}$ cm$^{-3}$) and in fact, it can be stated that a safe detection limit around 440 nm by such transmission experiments is ~ $10^{18}$ atoms of Fe per cm$^3$. Such a low absorption coefficient turns out to be lower than that of the orange crystals grown in [23] and [24], which exhibit $\alpha_{abs}$(< 550 nm) $\geq$ 2.5 cm$^{-1}$. The refractive index at 650 nm, obtained from $1+(1-T^2)^{1/2}/T$ (no Fresnel losses), is $\approx$ 1.96, close to the value ~ 1.91 given in [25] and to the values 1.87 – 2.01 (taking into account biaxiality of the material) obtained in [15] for the wavelengths 406 – 655 nm.

### 3.2 Luminescence under X-ray excitation

The luminescence of the $ZnMoO_4$ crystal sample (10 × 10 × 2 mm) was investigated as a function of temperature between 8 and 290 K under X-ray excitation. The sample was irradiated by X-rays from a BHV7 tube with a rhenium anode (20 kV, 20 mA). Light from the crystal was detected in the visible region by a FEU-106 photomultiplier (sensitive in the wide wavelength region of 300−800 nm) and in the near infrared region by FEU-83 photomultiplier with enhanced sensitivity up to $\approx$ 1 µm. Spectral measurements were carried out using a high-aperture MDR-2 monochromator. Emission spectra measured at 8, 118 and 290 K are shown in Fig. 2. A broad band in the visible region with a maximum at 550 nm was observed at room temperature. At 8 K luminescence exhibits an emission band with a maximum at $\approx$ 600 nm in agreement with the results of previous studies [15, 23, 26]. We have also observed a band at approximately 480 nm occurring below the liquid nitrogen temperature.

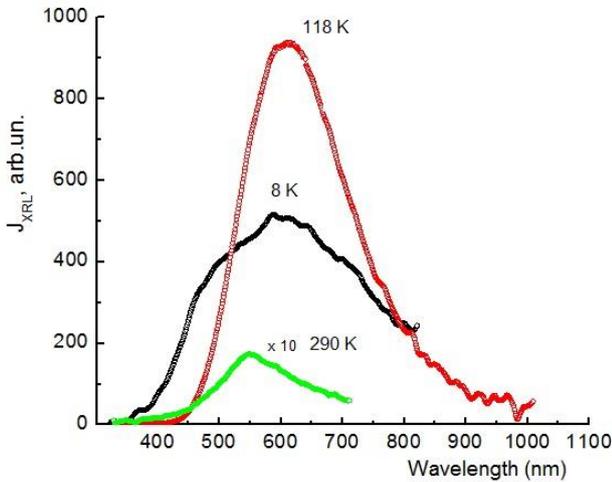

**Figure 2.** Emission spectra of $ZnMoO_4$ crystal under X-ray excitation at the temperatures 290, 118 and 8 K. $J_{XRL}$ denotes intensity of luminescence.

### 3.3 Magnetic susceptibility

Magnetic susceptibility was measured using a Quantum Design SQUID MPMS XL magnetometer operating in the 4.2 − 350 K temperature range and in the 0 − 5 T magnetic field range. The crystal mass was 210 mg and its volumic mass assumed to be 4.19 g/cm$^3$. It was mounted in a capsule placed in a straw and the negligibly small diamagnetic contribution ($|\chi| < 10^{-7}$) of the capsule was not subtracted from our data. The $ZnMoO_4$ proved to be weakly diamagnetic with a MKSA $\chi = -(8.0 \pm 0.2) \times 10^{-6}$ over the whole temperature range investigated, from 20 to 320 K. Thus, paramagnetic impurities such as $Fe^{2+}$ or $Fe^{3+}$ could not be evidenced even under higher applied magnetic fields up to 0.2 T.

### 3.4 Specific heat measurements

Specific heat measurements were made on a 3 × 3 × 2 mm$^3$ single crystal to optimize the exchange surface and avoid too much thermal inertia. The crystal was fixed on a sapphire sample holder with vacuum grease. The sample holder was mounted on the measurement shaft of a Quantum Design PPMS equipment interfaced to operate with a 2-$\tau$ pulse-step method corrected for the grease baseline. The results of the measurements are presented in Fig. 3. The phononic contribution could be approximated for temperatures higher than ~ 23 K by means of high-temperature series expansion:

$$C_{p,ph.} \propto 1 + \sum_{i=1}^{4} B_i \left[ 1 + \left( 2\pi \frac{T}{\theta_D} \right)^2 \right]^{-i} \quad (3)$$

(formula (5) from [27], and red curve in Fig. 3), which yielded a high Debye temperature of $\approx$ 625.1 K and the following Bernoulli numbers: $B_1$ = 1.9091, $B_2$ = 1.86714, $B_3$ = -0.96009, $B_4$ = -0.00907. No long range order (LRO) effect was observed down to 4 K. The $C_p$/3NR ratio at 351 K reaches 0.87 and remains lower than the Dulong and Petit limit, which suggests low anharmonic effects at play at this temperature, consistent with the high Debye temperature obtained by the HTS fit.

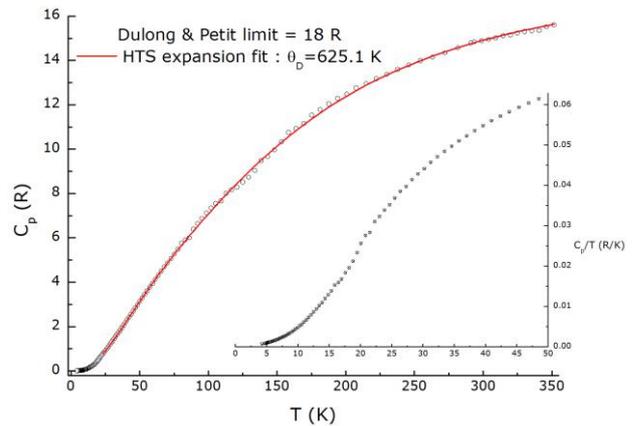

**Figure 3.** Specific heat *versus* temperature of a $ZnMoO_4$ single crystal. The inset shows $C_p$/T *vs* T at low temperature to evidence the absence of any LRO down to 4 K.

### 3.5 Low temperature tests

The operation of the scintillating bolometers with the produced $ZnMoO_4$ crystals at low temperatures was performed in the cryogenic laboratories of the CSNSM (Orsay). Two $ZnMoO_4$ samples (55 g and 160 g) were installed in a high-power dilution refrigerator with a large experimental space. A single photodetector, consisting of a 2" Ge disk and instrumented with a Neutron Transmutation Doped Ge thermistor as a temperature sensor, identical to those attached at the $ZnMoO_4$ crystals, collected the scintillation light emitted by both samples (see Fig. 4).

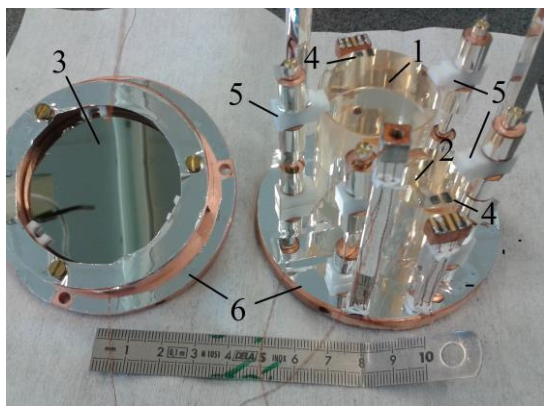

**Figure 4.** Photograph of the detectors setup: (1) $ZnMoO_4$ crystal $\varnothing 35 \times 40$ mm; (2) $ZnMoO_4$ crystal $\varnothing 20 \times 40$ mm; (3) Ge slab (photodetector); (4) NTD thermistors; (5) PTFE supporting elements; (6) Copper support of the detector covered by light reflector foil.

The copper heat-sink temperature was stabilized at 18 mK, and the detector operation temperature was about 1 mK higher due to sensor biasing. Both samples and light detector performed well, with an excellent signal-to-noise ratio. Unfortunately, aboveground operation is marginally compatible with such large crystals. In the large detector, practically every time window containing a full pulse contains also other pulses. One can see the pile-up effect on Inset of Fig. 5 where 4 seconds streaming data accumulated with a weak $^{232}$Th gamma source are shown. It should be stressed, the pile-up remains substantial also for the background data acquired without calibration source. This affects the energy resolution of the detector, which is expected to be much better in underground operation under heavy shielding with much lower pile-up effect.

In spite of that, a preliminary useful characterization can be performed in terms of signal amplitude, light yield, light quenching factors for alpha particles and crystal radiopurity. The results for the two crystals are summarized in Table 3. The difference in thermal response is due to the intrinsic irreproducibility of the thermal coupling in this type of detectors. The values of the quenching factor and of the light yield are perfectly compatible with those reported in the literature. Fig. 5 shows a $^{232}$Th calibration and Fig. 6 presents a light-heat scatter plot accumulated with the large $ZnMoO_4$ crystal. In terms of radiopurity, we will give quantitative estimations in a work in preparation. We can anticipate however that, with the exception of the $^{210}$Po line at 5.41 MeV, no internal alpha line emerged in the energy spectrum after about two weeks of data taking.

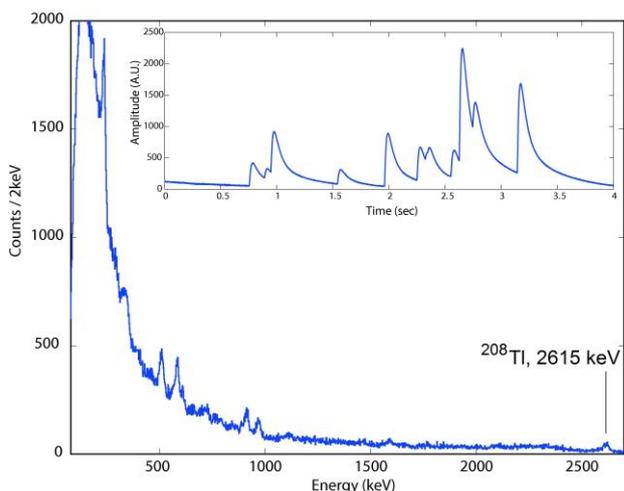

**Figure 5.** Energy spectrum accumulated by 160 g $ZnMoO_4$ bolometer with $^{232}$Th γ source. (Inset) Typical pile-up effect due to a slow time response of the bolometric detector.

**Table 3.** Performance of 55 g and 160 g $ZnMoO_4$ detectors. QF denote quenching factor of α particle signals with respect to β particle signals for the same deposited energy (~ 5.4 MeV).

| Parameter | 55 g | 160 g |
|---|---|---|
| Light yield (keV/MeV) | 0.98 | 0.96 |
| QF | 0.153 | 0.156 |

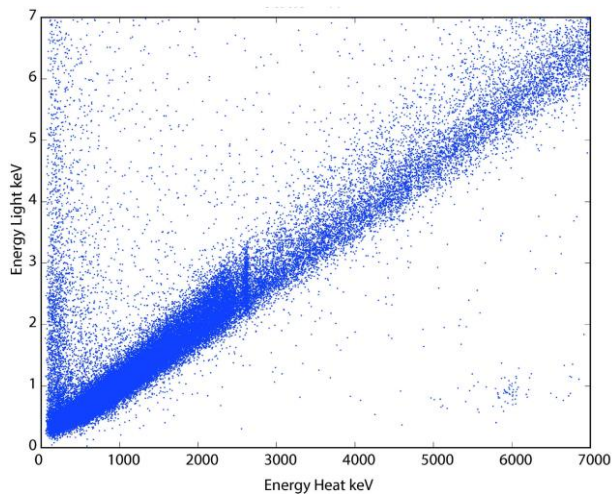

**Figure 6.** The scatter plot of light and heat signals for 160 g $ZnMoO_4$ crystal accumulated over 81 h in aboveground set-up in Orsay. The α band (populated mainly by $^{210}$Po internal contamination of the crystal) is neatly separated from the β band (containing also γ quanta and cosmic muons).

## 4 Conclusions

The LUMINEU program aims at performing a pilot experiment on neutrinoless double beta decay of $^{100}$Mo using radiopure $ZnMoO_4$ crystals operated as scintillating bolometers. This problem requires development of methods of molybdenum purification to obtain crystals with desired characteristics.

Different approaches of molybdenum purification for $ZnMoO_4$ crystals growth were elaborated. A purification using two stages sublimation (with addition of zinc molybdate) and recrystallization from aqueous solutions of ammonium para-molybdate (using zinc molybdate as a collector) is a promising approach to purify molybdenum for high quality radiopure $ZnMoO_4$ crystals growth.

A first batch of LUMINEU crystals with mass up to 1.5 kg have been successfully grown by by the low-thermal-gradient Czochralski technique, and their optical, luminescent, diamagnetic, thermal and bolometric properties were tested. Characterization of the material is in progress.

In the future, crystals of increasing mass from deep purified precursors will be developed for the LUMINEU experiment, including crystals enriched in the isotope $^{100}$Mo.

## Acknowledgments

The development of $ZnMoO_4$ scintillating bolometers is part of the LUMINEU program (Luminescent Underground Molybdenum Investigation for NEUtrino mass and nature), a project receiving funds from the L'Agence nationale de la recherche (France). The work was supported in part by the project "Cryogenic detector to search for neutrinoless double beta decay of molybdenum" in the framework of the Programme "Dnipro" based on Ukraine-France Agreement on Cultural, Scientific and Technological Cooperation.